\documentclass[aip,graphicx]{revtex4-1}
\usepackage{graphicx}
\usepackage{siunitx}
\usepackage{soul}

\usepackage[english]{babel}
\usepackage{blindtext}

\newcommand{\LAO}{LaAlO\textsubscript{3}\xspace}
\newcommand{\Fe}{Fe\textsubscript{3}O\textsubscript{4}\xspace}

\newcommand{\STO}{SrTiO\textsubscript{3}\xspace}

\usepackage{xcolor}
\usepackage{lineno}
\usepackage{amsmath}

\draft 

\begin{document}

%\pagestyle{fancy}
%\rhead{\includegraphics[width=2.5cm]{vch-logo.png}}

%\linenumbers

%%%%%%%%%%%%%%%%%%%%%%%%%

\title{Tunable 2D Electron- and 2D Hole States Observed at Fe/\STO Interfaces}

%\maketitle

\author{Pia M. D\"uring}
\affiliation{Fachbereich Physik, Universit\"at Konstanz, 78457 Konstanz, Germany}

\author{Paul Rosenberger}
\affiliation{Fachbereich Physik, Universit\"at Konstanz, 78457 Konstanz, Germany}
\affiliation{Fakult\"at Physik, Technische Universit\"at Dortmund, 44221 Dortmund, Germany}

\author{Lutz Baumgarten}
\affiliation{Forschungszentrum Jülich GmbH, Peter Grünberg Institut (PGI-6), 52425 Jülich, Germany
}

\author{Fatima Alarab}
\affiliation{Paul Scherrer Institute, Swiss Light Source, CH-5232 Villingen PSI, Switzerland
}

\author{Frank Lechermann}
\affiliation{Institut f\"ur Theoretische Physik III, Ruhr-Universit\"at Bochum, 44780 Bochum, Germany
}

\author{Vladimir N. Strocov}
\affiliation{Paul Scherrer Institute, Swiss Light Source, CH-5232 Villingen PSI, Switzerland
}

\author{Martina M\"uller}
\email{\\
\textbf{Corresponding author:} martina.mueller@uni-konstanz.de}
\affiliation{Fachbereich Physik, Universit\"at Konstanz, 78457 Konstanz, Germany}

% Affiliations: Please provide adacemic titles (Prof. or Dr.) for all authors where applicable, and include an institutional email address for all corresponding authors
%\begin{affiliations}
%P. M. D\"uring, P. Rosenberger, Prof. M. M\"uller \\
%Fachbereich Physik, Universit\"at Konstanz, 78457 Konstanz, Germany \\
%Corresponding author: martina.mueller@uni-konstanz.de \\

%P. Rosenberger \\
%Fakult\"at Physik, Technische Universit\"at Dortmund, 44221 Dortmund, Germany

%Dr. L. Baumgarten \\
%Forschungszentrum Jülich GmbH, Peter Grünberg Institut (PGI-6), 52425 Jülich, Germany

%Dr. F. Alarab, Dr. V. N. Strocov \\
%Paul Scherrer Institute, Swiss Light Source, CH-5232 Villingen PSI, Switzerland

%Dr. F. Lechermann \\
%Institut f\"ur Theoretische Physik III, Ruhr-Universit\"at Bochum, 44780 Bochum, Germany

%\end{affiliations}

%%%%%%


% 200 Words!!
\begin{abstract}
Oxide electronics provide the key concepts and materials for enhancing silicon-based semiconductor technologies with novel functionalities. However, a basic but key property of semiconductor devices still needs to be unveiled in its oxidic counterparts: the ability to set or even switch between two types of carriers  - either negatively (n) charged electrons or positively (p) charged holes.  Here, we provide direct evidence for individually emerging n- or p-type 2D band dispersions in STO-based heterostructures using resonant photoelectron spectroscopy. The key to tuning the carrier character is the oxidation state of an adjacent Fe-based interface layer: For Fe and FeO, hole bands emerge in the empty band gap region of STO due to hybridization of Ti- and Fe- derived states across the interface, while for \Fe overlayers, an 2D electron system is formed. Unexpected oxygen vacancy characteristics arise for the hole-type interfaces, which as of yet had been exclusively assigned to the emergence of 2DESs. In general, this finding opens up the possibility to straightforwardly switch the type of conductivity at STO interfaces by the oxidation state of a redox overlayer. This will extend the spectrum of phenomena in oxide electronics, including the realization of combined n/p-type all-oxide transistors or logic gates.
\end{abstract}

\maketitle

%%%%%%%%%%%%%%%%%%%%%%%%%%%%%%%

% Text: Please use section headings and subheadings as specified below. For communications, all section headings apart from Experimental Section should be removed
% Please make the first reference to a display item bold: \textbf{Figure 1}
% Do not abbreviate Figure, Equation, etc.; display items are always singular, i.e., Figure 1 and 2.
% Equations are always singular, i.e., Equation 1 and 2, and should be inserted using the {equation} environment, not as graphics
% Please do not use footnotes in the text, additional information can be added to the Reference list.

\subsection*{Introduction}

Conductivity in electronic materials originates either from negatively (n) charged electrons or positively (p) charged holes. When n- or p-type carriers are confined at interfaces between semiconductors or insulators, either a two-dimensional electron system (2DES) or -hole system (2DHS) can form, providing the physical grounds for all transistor-type devices and quantum wells exploited in semiconductor (SC)  technology. Metal-oxide SCs exhibit more versatile properties than silicon or metal-nitride SCs, such as the coexistence of ferromagnetism and superconductivity\cite{Reyen2007}, large spin splitting, and large magnetoresistance\cite{Brinkman2007,Lee2013} - if it were not for the fact that there are many n-type but very few p-type metal-oxide materials. 

The existence of a 2DES was first observed at the oxide-oxide interface of the now classic \LAO(LAO)/\STO(STO) heterostructure\cite{Ohtomo2004}, and to date at numerous other STO-based systems \cite{}. More recently, a 2DHS was found in a STO/LAO/STO trilayer structure - but coexisting with a spatially separated 2DEG, making individual addressing of hole-related properties difficult\cite{Lee2018}. Perfectly sharp and crystalline ordered interfaces were necessary to stabilize a 2DES and combined 2DHS/2DEG in the \LAO(LAO)/\STO(STO)-based heterostructures\cite{Ohtomo2004,Lee2018}, and to suppress the formation of oxygen vacancies (V$_{\mathrm{O}}$). 

Individual 2D hole-type conductivity at metal oxide interfaces was recently revealed in amorphous and air-exposed AlO$_x$/Fe$_x$O$_y$/STO systems by electrical transport experiments, reporting high mobilities up to 24.000 cm$^2$ V$^{-1}$ s$^{-1}$ \,\cite{Anh2020}. A sign change in the Hall voltage was interpreted as a switching from mainly hole to mainly electron-type carrier conduction above an Fe$_x$O$_y$ threshold thickness of \SI{0.20}{\nano\meter}. The transition from p- to n-type conduction was associated with an increase of oxygen vacancies at the \STO interface due to the scavenging of oxygen with increasing Fe metal coverage, while p-type carriers were assigned to Fe-related cation hole doping in the \STO \cite{Chambers2015}. 

The respective physical origins of 2DES and 2DHS emerging at STO-based interfaces are still subject of intense debate. In the case of 2DES, models describe very different scenarios that are result of either polar catastrophe\cite{Nakagawa2006}, formation of oxygen vacancies\cite{}, quantum confinement\cite{} or cationic mixing at the interface\cite{}. 
In particular, the reduction of Ti$^{4+}$ to Ti$^{3+/2+}$ on the \STO side of the interface and concomitant release of itinerant electrons is assumed to be the origin for the formation of 2DEGs when oxygen vacancies are formed. Experimentally, 2D band dispersions of free electrons have been observed at x-ray irradiated bare STO surfaces\cite{Santander-Syro2011} and at redox-created STO interfaces, e.g. Al/STO\cite{Roedel2016} or Eu/STO\cite{Loemker2017,Loemker2019}.  However, direct evidence for the emergence of 2D hole character in terms of band dispersion at STO-based interfaces and knowledge of the corresponding Ti$^{3+/2+}$ interface chemistry remains elusive to date, why modelling is still pending.

%%%%%%%%%%%%%%%%%%%%%%%%%%%%%%%

Here, we demonstrate the emergence of individual, i.e. either hole- or electron-type band dispersions at Fe-based \STO interfaces using angle-resolved photoelectron spectroscopy (ARPES) as a direct probe of the momentum-resolved occupied electronic band structure $E(\mathbf{k})$. The p- and n-type interfaces are generated by depositing ultrathin Fe~(2 unit cells (uc)), FeO~(2\,uc) and Fe$_3$O$_4$~(1\,uc) layers on TiO$_2$- terminated \STO~(100) substrates, see Experimental Section. We assign the corresponding Ti$^{3+/2+}$ interface chemistry determined by angle-integrated x-ray photoelectron spectroscopy (XPS) on as-grown, pristine and x-ray exposed Fe-based \STO interfaces.

\begin{figure*}[htb]
\includegraphics[width=1\textwidth]{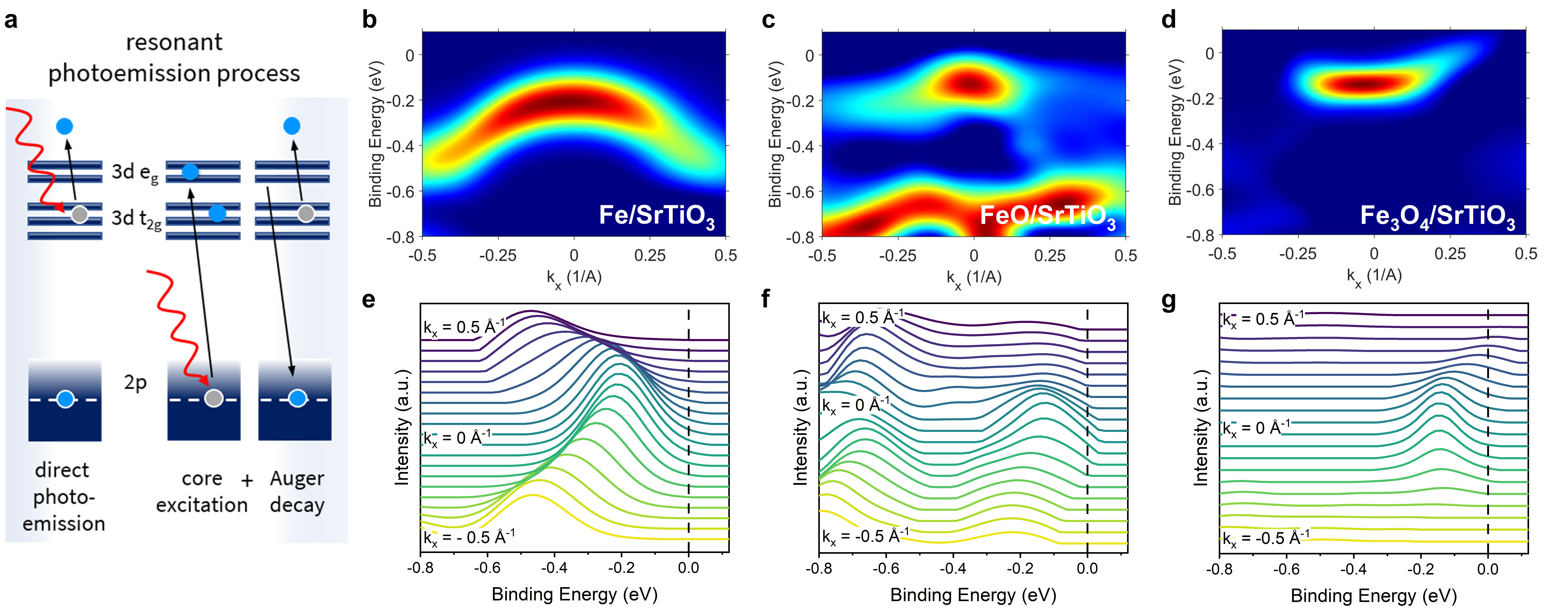}
\caption{\textbf{Energy dispersion maps of p- and n-type valence bands emerging in the otherwise empty band gap at Fe--\STO interfaces.} \textbf{a}, Schematics of the resonant soft x-ray photoemission process boosting the spectroscopic contrast for the Ti $3d$-derived states at the buried Fe--\STO interfaces. \textbf{b-d}, Energy dispersion maps $E(\mathbf{k})$ of the valence bands just below the Fermi level for \textbf{b}) Fe/STO, \textbf{c}) FeO/STO and \textbf{d}) \Fe/STO interfaces. The photon excitation energy was set to the resonant Ti\,$L_{3/2}$ absorption edge at $h\nu$ = 466\,eV using circular- (c+) and s-polarized x-rays, thereby capturing $d_{xz}$, $d_{yz}$ and $d_{xz}$ (c+ pol.) as well as $d_{yz}$ and $d_{xz}$ (s-pol.) bands, respectively. Depending on the particular Fe--\STO interface, either hole- or electron band dispersions emerge in the otherwise unoccupied band gap of STO. The intense resonant behaviour of these gap states demonstrates their high degree of localization on the Ti atoms. The emergence of either hole- or electron band character depends on the specific interface chemistry of the Fe--\STO heterostructures. \textbf{e-g}, Corresponding energy distribution curves (EDCs) after background correction, see Experimental Section.}%
\label{fig:HighStats}
\end{figure*}

Soft x-ray excitation of photoelectrons increases the probing depth compared to conventional surface sensitive ultraviolet ARPES to several nanometers~(nm), and thereby enables access to the electronic structure of buried interfaces\cite{Cancellieri2018}. A resonant enhancement of the Ti- and Fe- related valence band (VB) intensity is accomplished by scanning the photon energy across the Ti\,$L_{3/2}$ (\SI{466}{\electronvolt}) and Fe\,$L_{3/2}$ (\SI{710}{\electronvolt}) absorption edges, respectively. The resonant photoemission process is schematically depicted in Figure\,\ref{fig:HighStats}a. It causes an interference of the direct photoemission of a VB electron with the coincident two-step autoionization (Auger) process into the same final state, where, at the $L$-edges, the Auger contribution boosts the Ti/Fe signals\cite{Strocov2010,Cancellieri2018}.

%%%%%%%%%%%%%%%%%%%%%

\subsection*{p- and n-type \STO interfaces: Experiment and Theory}

Figure~\ref{fig:HighStats}~b--d depict the energy dispersion $E(\mathbf{k})$ of the valence bands right below the Fermi level $E_F$  around the $\Gamma$ point of the Brillouin zone (BZ). Whereas for bare STO substrates no states exist in the band gap, the photoemission spectra of all Fe--\STO interfaces show up dispersive Ti-related features, for which the spectroscopic contrast is resonantly enhanced. For the Fe/STO~(b) and FeO/STO~(c) interfaces, the valence bands bend parabolically downwards from a maximum centered at the zone centre $\Gamma$. This feature clearly indicates the existence of hole states with negative effective mass $\delta^2 E/\delta k^2 <0$. At the \Fe/STO interface (d), in contrast, a dispersion with positive effective mass is observed at $E_F$, which is a hallmark for a 2D electron-like system. 

The corresponding energy distribution curves (EDCs) are shown in Figure~\ref{fig:HighStats}~e--g.  Details on the background correction are given in Experimental Section. We determine the hole band maxima at approximately \SI{-200}{\milli\electronvolt} (Fe/STO) and \SI{-110}{\milli\electronvolt} (FeO/STO) below $E_F$, while the minimum of the confined electronic bands at the \Fe/STO interface is located at \SI{-120}{\milli\electronvolt}. The lower binding energy of the p-band at the Fe/STO interface represents a higher hole energy compared to FeO/STO, but without cutting the Fermi level. For the FeO/STO sample, the hole band shifts upwards to $E_F$. From the EDC in Figure~\ref{fig:HighStats}~f, we infer that the p-type hole band is closely approaching the Fermi level at the zone centre $\Gamma$. Furthermore, the dispersive feature at the \Fe/STO  interface crosses $E_F$ at $k_x \sim \pm\SI{0.3}{\angstrom}^{-1}$ and thus evidently represents a 2DES.
%Both c+ and s-polarizations were choosen for technical reasons only, all relevant spectral information is contained in either case. 

\begin{figure*}[htb]
\includegraphics[width=1\textwidth]{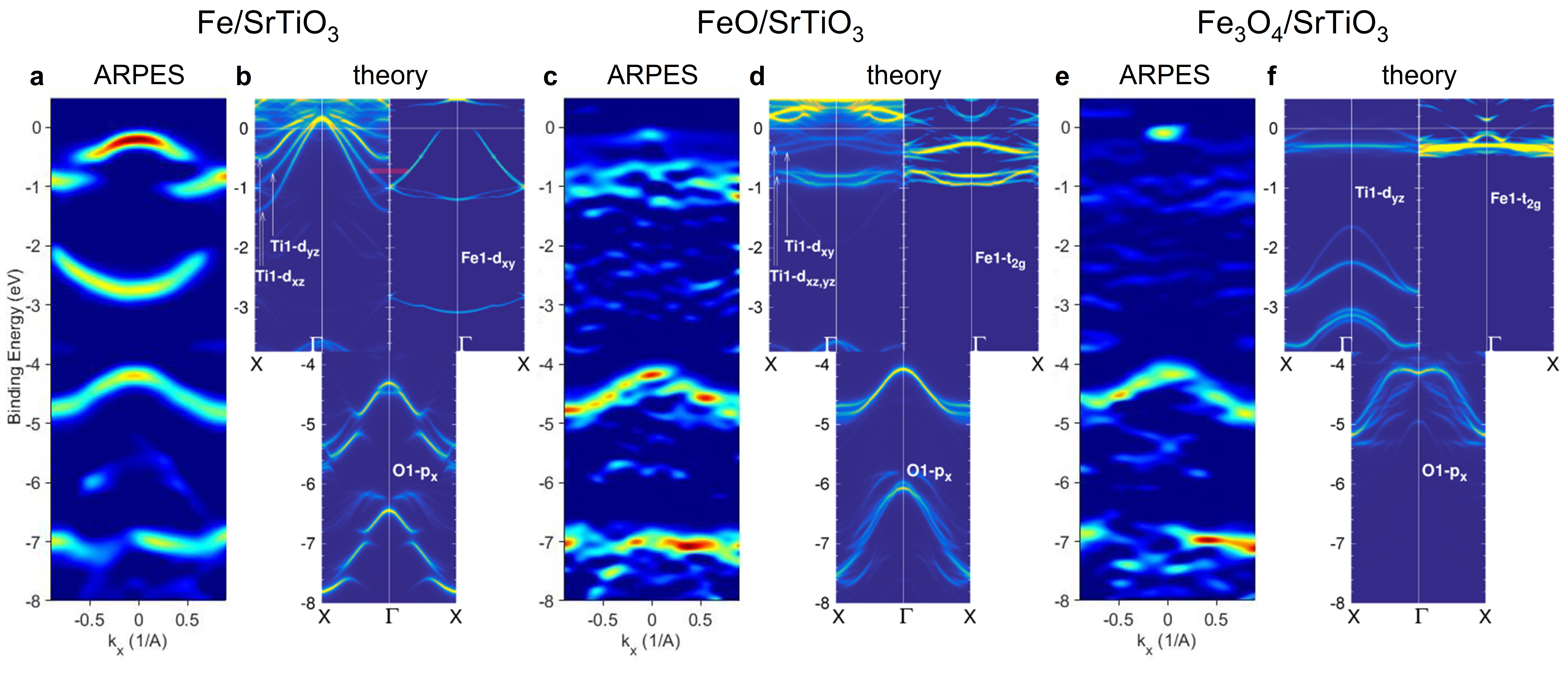}
\caption{\textbf{Experimental ARPES band structure and supercell calculations for Fe/STO, FeO/STO and Fe$_3$O$_4$/STO interfaces.} \textbf{a,\,c,\,e}, Experimental band dispersions $E(\mathbf{k})$ along X$-\Gamma-$X for \textbf{a}) Fe/STO, \textbf{c}) FeO/STO and \textbf{e}) \Fe/STO interfaces, measured at the Ti$^{4+}$ $t_{2g}$ resonance condition ($h\nu$ = 466\,eV) using c+ polarization. \textbf{b,\,d,\,f}, Corresponding first-principles dispersions obtained from projecting onto Ti, Fe and O sites with $d$ or $p$ orbital character in the respective first ('1') interface layer. Ti1 (Fe1) projections are shown in the left (right) panels for the energy window $[-3.75,0.5]$\,eV and O1 projection in the bottom panels for the energy window $[-8,-3.75]$\,eV. The red bar in \textbf{b} marks the low-energy hybridizing Ti1 $d_{yz}$ and Fe1 $d_{xy}$ branches. With increasing Fe oxidation state, the hole character of the dominant Ti1 $d_{xz,yz}$ dispersions systematically diminish their spectral weight.}
\label{fig:DFT}
\end{figure*}

In order to capture the entire valence electronic structure of the Fe, Ti and O subsystems, we show the extended ARPES images within the energy window from $-8$ to $0$\,eV in Figure~\ref{fig:DFT}~a,\,c,\,e. The experimental energy dispersions $E(\mathbf{k})$ along X$-\Gamma-$X in the BZ are depicted for a) Fe/STO, c) FeO/STO and e) Fe$_3$O$_4$/STO. In addition to the near-$E_F$ region, as was shown in Figure~\ref{fig:HighStats}\,b-d, here also the energetically lower Fe~$3d$ and O~$2p$ VBs become discernable. By using circularly (c+) polarized x-rays, the full manifold of the Ti\,$3d$ orbital-related band symmetries ($d_{xz}$, $d_{yz}$ and $d_{xz}$) was recorded. % for the Fe- and FeO/STO samples. For \Fe/STO, s-polarized x-rays map the $d_{yz}$ and $d_{xz}$ bands. 

A juxtaposition of the experimental ARPES results to first-principles supercell (SC) band structure calculations is given in Figure~\ref{fig:DFT}~b,\,d,\,f. The theoretical dispersions were obtained from projecting onto the Ti, Fe and O lattice sites and orbital characters ($d$ and $p$) in the first layers adjacent to the interface (index '1'), see Supporting Information Figure~\ref{fig:struc}. Here, we show the Ti1 (Fe1) projection on the left (right) panels for the energy window $[-3.75,0.5]$\,eV, and O1 projection (bottom panels) for the energy window $[-8,-3.75]$\,eV. We note that the theoretical intensity of the Ti projections is multiplied by five to increase visibility, and low-intensity weight from projecting onto Fe states has been cut off. This procedure seems reasonable with regard to the experimental resonance condition enhancing the Ti contribution in the valence band. For the theoretical \Fe/STO case in Figure~\ref{fig:DFT}~f, the single O1 projection misses the high-energy weight below $-6$~eV, because of inequivalent Fe/O sites of the inverse spinel lattice model. Extending the projection to additional O sites recovers that weight, but we have for brevity ignored this point.

The experimental dispersion features (Figure~\ref{fig:DFT}~a,\,c,\,e) are in good accordance with the band structure calculations (Figure~\ref{fig:DFT}~b,\,d,\,f) for the given Fe--\STO interfaces. Especially the hole (p-type) character of the Fe/STO interface is well reproduced. Remarkably, above the STO O~$2p$ VB, from $\sim\SI{3.75}{\electronvolt}$ up to the Fermi level, the otherwise empty band gap of \STO is filled by a manifold of Ti~$3d$-derived states. Experimentally, in this energy range the Fe/STO interface is strongly dominated by a non-dispersive Fe~$3d$ density of states, which is resonating at the Fe $L_{3/2}$ but not at the Ti $L_{3/2}$ edge (see Supporting Information, Figure~\ref{fig:ResPESTi}~a,\,b). The best agreement between SC calculations and experiment is achieved with regard to the leading Fe $3d$ orbital sectors with strongest spectral weight. 

Additionally, the relevant Ti~$3d$ low-energy spectral weight is generally smaller than the Fe~$3d$ one, such that the effective Fe-Ti hybridization appears to be the main driving force behind the observed confined hole states. A further general key observation from theory is that only the directly adjacent TiO$_2$ and Fe(O) interface layers (here termed with index '1', respectively) need to be considered, since further layers more distant to the interface are not decisive for the hole vs. electron characteristics. In particular on the relevant STO side, the hole state is confined solely within this first TiO$_2$ layer (or 1\,uc), which is apparently different to e.g. the more extended electron gas in LAO/STO. Among all Ti~$3d$ states, the spectral weight of Ti~$d_{xz,yz}$ orbitals dominates via energy-dependent hybridization to Fe~$t_{2g}$ over that of in-plane Ti~$d_{xy}$ states. For instance for the Fe/STO case in Figure~\ref{fig:DFT}~b, the hole dispersion of Ti1-$d_{yz}$ hybridizes with Fe1 $d_{xy}$ (see red bar), and is thus well observable in the ARPES spectra. On the other hand, the Ti $3d$ weight on the deeper lying Fe1 $d_{xy}$ dispersion between $[-3,-2]$\,eV is much weaker. With regard to the hole character, we find the dominant Ti1 $d_{xz,yz}$ parabolic bands systematically diminishing their spectral weight with increasing Fe oxidation state (see energy window $[-0.75,0]$\,eV of left panels in Figure~\ref{fig:DFT}~b,\,d,\,f). This theoretical trend is fully in line with the experimental results in Figure~\ref{fig:DFT}~a,\,c,\,e. We note that $E_F$ may be different for the real systems presumably including oxygen vacancies, see below.

%Concerning the evolution of low-energy carriers with Fe-based system, a diminishing of the hole-character strength is observed from focussing on the dominant Ti1 $d_{xz,yz}$ dispersion in the energy window $[-0.75,0]$\,eV (see left panel in Fig.~\ref{fig:DFT}b,d,f), which is in line with the experimental trend for the real systems (presumably including oxygen vacancies, see below).

%%%%%%%%%%%%%%%

\subsection*{Interface chemical states and driven redox processes}

\begin{figure*}[htb]
\includegraphics[width=1\textwidth]{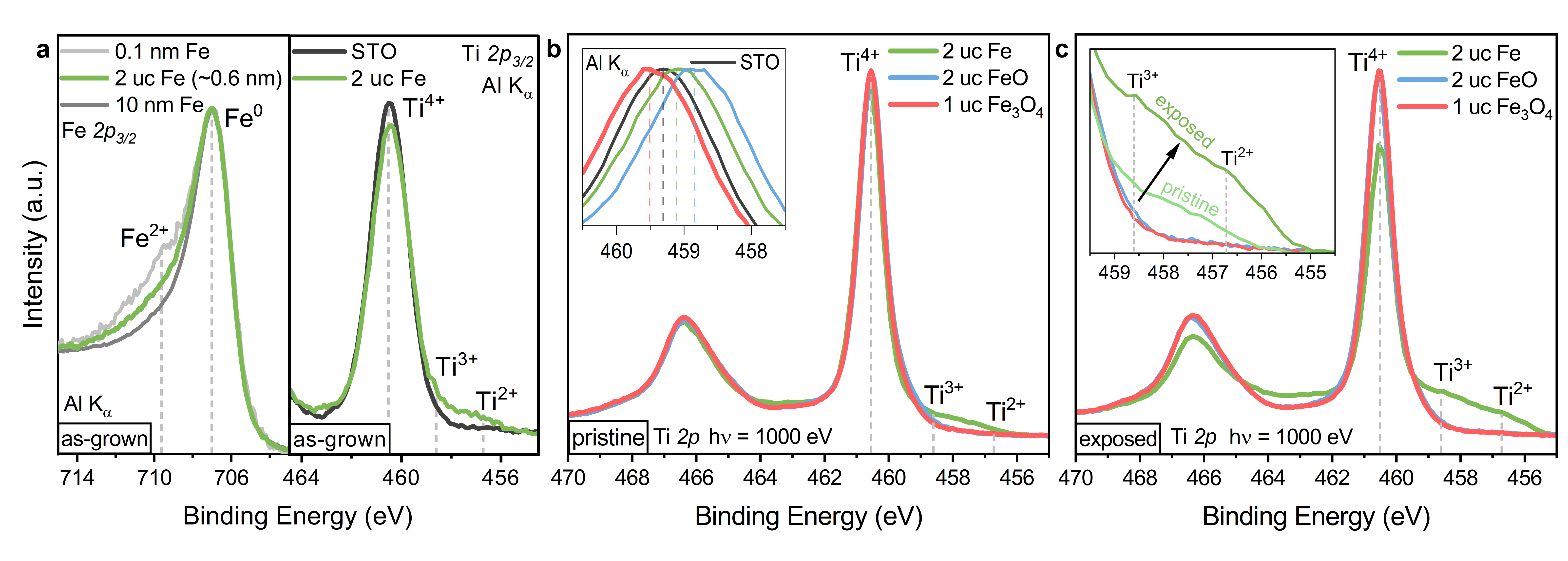}
\caption{\textbf{Interface chemical properties of Fe--\STO samples in the as-grown, pristine and x-ray exposed states.} \textbf{a} XPS data of the as-grown Fe/\STO sample recorded at RT. (left panel) Fe\,$2p_{3/2}$ core-level for three different Fe coverages indicating an interfacial Fe$^{2+}$ compound. (right panel) Ti\,$2p_{3/2}$ core-levels of the \STO substrate before and after Fe deposition, revealing a Ti$^{3+/2+}$ component after Fe deposition.  
\textbf{b,\,c} Ti\,$2p_{3/2}$ and $2p_{1/2}$ core-levels of \textbf{b} pristine and \textbf{c} x-ray exposed Fe/STO, FeO/STO and \Fe/STO interfaces recorded at 12\,K. No sizeable amount of Ti$^{3+,2+}$ is discerned for the FeO/\STO and \Fe/\STO samples in both states, while x-ray exposure further increases the Ti$^{3+,2+}$ weight at Fe/\STO interface (inset \textbf{c}). All Ti $2p$ peaks are shifted and normalized for better comparison. Inset \textbf{b} shows the non-shifted, as measured energy positions of the Ti\,$2p_{3/2}$ peaks, from which the band alignment is estimated. XPS spectra of as-grown samples were collected using Al K$_{\alpha}$ radiation ($1486.6$\,eV), while the pristine and exposed states were recorded at the synchrotron using $h\nu$ = 1000\,eV.}%
\label{fig:XPS}
\end{figure*}

Obviously, the emergence of either hole- or electron band dispersions is closely linked to the specific chemical bonding at the Fe--\STO interfaces. As we shall see in the following, it is reasonable to discriminate between growth- and x-ray-induced properties. To this end, XPS experiments were performed at three points in time; i) in the laboratory on the 'as-grown' samples, ii) at the synchrotron using the intense x-ray beam  on a fresh spot on the sample ('pristine') and iii) after $>$\SI{15}{\minute} illumination ('exposed'). Following the element-specific chemical binding energy (BE) shifts as well as the rigid BE shifts of all core-levels peaks allows to identify charge transfer, redox processes and band bending at the Fe--\STO interfaces, respectively.

Figure~\ref{fig:XPS} compares the XPS spectra of p-type Fe/STO, FeO/STO and n-type \Fe/STO samples in their 'as grown', 'pristine' and 'exposed' states. As becomes apparent from a visual inspection of the Ti\,$2p$ core-level line shapes, the p-type Fe/STO interface undergoes a redistribution of spectral weight -- best observed on the lower binding energy side of the Ti\,$2p_{3/2}$ peak -- first during growth (Figure~\ref{fig:XPS}~a) and then during x-ray exposure (Figure~\ref{fig:XPS}~b,\,c). In contrast, the n-type \Fe/STO and p-type FeO/STO samples remains almost unchanged. The reduction of Ti, which formally is associated with the transformation of Ti$^{4+}$ to Ti$^{3+/2+}$ ions, results in a release of electrons. These excess negative charges are typically accompanied by the formation of single or even clustered oxygen vacancies $V_{\mathrm{O}}$ (in the case of Ti$^{2+}$) at the \STO side of the interface, and screen these positively charged defects. Therefore, an intensity redistribution from the main Ti$^{4+}$-related peak to the Ti$^{3+/2+}$-related shoulder is commonly associated with the emergence of an n-type 2D electron system -- but here is unexpectedly observed for the p-type Fe/STO interface. Vice versa, its almost absence in case of the n-type \Fe/STO sample is astonishing, too. 

For creating the p-type Fe/STO interface, pure Fe metal was deposited onto TiO$_2$-terminated STO(001) surfaces (see Experimental Section). Fe has a much weaker oxygen affinity in comparison to other metals, e.g. Al\cite{Roedel2016} or Eu\cite{Loemker2017, Loemker2019}, and therefore is significantly less prone to oxidation in contact with \STO. In the present case, a small fraction of about 10\% Fe$^{2+}$ spectral weight is discerned from the as-grown 2\,uc Fe/STO spectra (see Figure~\ref{fig:XPS}~a). The fractional oxidation from Fe$^0$ to Fe$^{2+}$ can be associated with two different scenarios; it may arise from (i) oxygen scavenging out of the \STO substrate and redox-reaction to FeO or (ii) diffusion of Fe into the STO topmost layer occupying Ti lattice sites\cite{Chambers2015}. A redox-reaction to FeO goes along with the formation of oxygen vacancies and the release of two excess electrons per $V_{\mathrm{O}}$. This process is detectable from the 'classical' Ti$^{4+}$ to Ti$^{3+/2+}$ reduction signatures in Ti $2p$ photoemission spectra. On the other hand, a cation lattice site substitution Fe$_{\mathrm{Ti}}$ causes the transfer of both $V_{\mathrm{O}}$ electrons to the Fe$_{\mathrm{Ti}}$ site, making it Fe$^{2+}$ valency\cite{Chambers2015}. In the last case, positively charged oxygen vacancies (holes) are created at the Fe/STO interface, but without a Ti$^{4+}\rightarrow \mathrm{Ti}^{3+}$ reduction and without release of excess electrons. Instead, the thus formed Fe$_x$Ti$_{1-x}$O$_{2-x}$ layer may cause an upward band shift of the STO ($\sim$\SI{0.23}{\milli\electronvolt})\cite{Chambers2015}, see below. 

A well-known phenomenon in the field of x-ray photoelectron spectroscopy on \STO is x-ray illumination creating a self-saturating amount of oxygen vacancies. Assorted experiments have demonstrated this effect for bare \STO surfaces as well as for buried interfaces\cite{Santander-Syro2011,Walker2015,Strocov2019}. Light-induced $V_{\mathrm{O}}$ creation is also assigned to the formation of a 2D electron system at the \STO interface, where the electron accumulation layer screens the $V_{\mathrm{O}}$ hole defects\cite{Dudy2016}. Comparing the Ti $2p$ core-level spectra in Figure~\ref{fig:XPS} for (b) pristine and (c) exposed samples reveals the following evolution of the Fe--\STO systems under x-ray illumination: For Fe/STO, we find a significant amount of Ti$^{3+,2+}$ emerging at the lower BE side of the Ti~$2p$ doublet, which increased after x-ray exposure from $\sim 7 \pm 2 \%$ ('pristine') to approximately $\sim 18 \pm2$\% ('exposed'). In stark contrast, for the FeO/STO and \Fe/STO interfaces the Ti~$2p$ line shapes remain almost completely unaffected by x-ray illumination, hence no x-ray induced chemical process and oxygen vacancy formation are discernable from the Ti $2p$ core-level signatures. We note that the \STO surfaces feature subtle experimental difference for subsequent Fe or FeO, \Fe coverage: Before Fe deposition, the TiO$_2$-terminated \STO surface is oxygen deficient due to annealing in vacuum, hence making it charge neutral\cite{Cheng2022}. Before FeO and \Fe growth, in contrast, surface oxygen vacancies were annihilated by introducing molecular oxygen, see Experimental Section. Therefore, the recovered TiO$_2$-terminated \STO surface is weakly polar\cite{Cheng2022}. To conclude on any light-induced oxygen release across the \STO interface and possible reduction or oxidation reaction with the overlaying Fe, FeO or \Fe layers, we examined the Fe\,$2p$ core-levels in the pristine and exposed states (shown in the SI), but no changes of the line shapes could be discerned for Fe/\STO, FeO/\STO and \Fe/\STO. Thus, the ARPES experiment unveils intrinsic hole-type interface states probably insensitive to x-ray illumination.

\begin{figure*}[h!]
\centering
\includegraphics[width=0.9\textwidth]{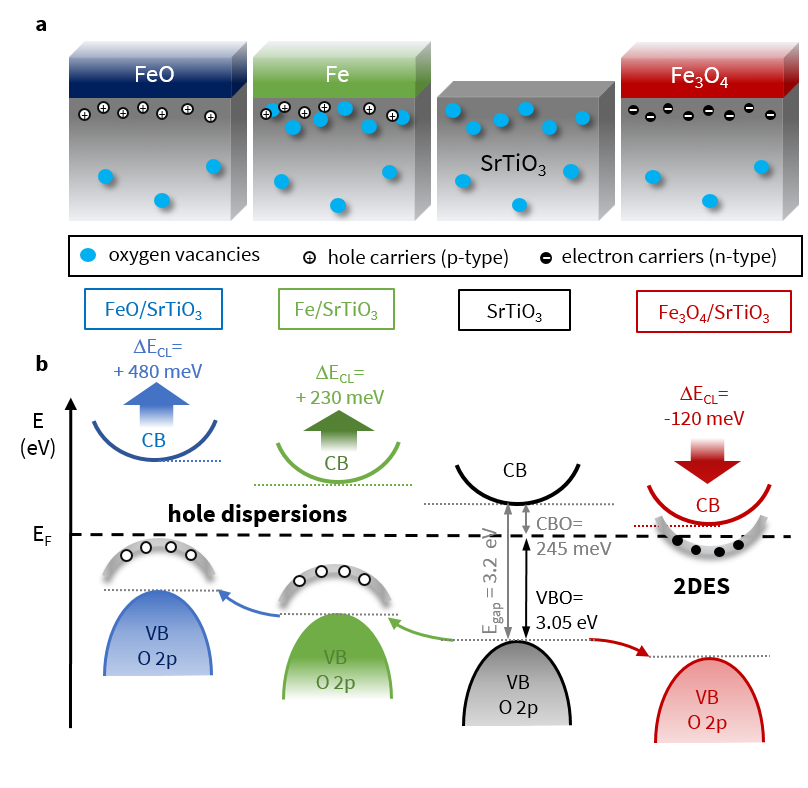}
\caption{\textbf{Experimentally derived interface formation and band alignment of Fe--\STO heterostructures for different oxidation states of the Fe-based overlayers.} \textbf{a} Schematics of the Fe--\STO interface formations as derived from XPS analysis. \textbf{b} Schematic band alignment as determined from XPS. The valence band maximum of a bare \STO substrate is located \SI{3.05}{\electronvolt} below $E_F$ (VB offset, VBO). Assuming an STO band gap of $E_{\mathrm{gap}}$ = \SI{3.2}{\electronvolt}, the unoccupied conduction band minimum is located \SI{245}{\milli\electronvolt} above the Fermi level (CB offset, CBO). From the rigid core-level shifts $\Delta E_{\mathrm{CL}}$, the respective band alignment was estimated for the of Fe/\STO ($+\SI{230}{\milli\electronvolt}$), FeO/\STO ($+\SI{480}{\milli\electronvolt}$) and \Fe/\STO ($-\SI{120}{\milli\electronvolt}$) interfaces. Whereas for Fe/\STO oxygen vacancies are likely present at the interface, for FeO/\STO and \Fe/\STO no experimental hints in terms of Ti$^{3+/2+}$ contributions could be discerned. }
\label{fig:model}
\end{figure*}

\subsection*{Band alignment at Fe--\STO interfaces}
Finally, the electrostatic interface properties need to be put into perspective. Both the presence of oxygen vacancies and concomitant screening by an electron accumulation layer as well as the polarity of adjacent intralayers change the interfacial band alignment (BA) compared to the uncovered, undoped STO surface. To this end, the evolution of the interfacial BA is deduced as a flat band model from the rigid core-level BE shifts of the 'as grown' sample set. For p-type Fe/STO and FeO/STO, the Ti~$2p$ core-level spectra (see inset in Figure~\ref{fig:XPS}~b) shift towards lower binding energy $\Delta E_{\mathrm{CL}}= \SI{-0.48}{\electronvolt} $ ($\Delta E_{\mathrm{CL}} =$ \SI{-0.23}{\electronvolt}), whereas a shifts to higher BE ($\Delta E_{\mathrm{CL}} =$ \SI{0.12}{\electronvolt}) is observed in the case of \Fe/STO. From the XPS valence band spectrum of a bare STO substrate, we determine the O~$2p$ valence band offset (VBO) at $3.05\pm0.08$\,eV below the Fermi level. Assuming a bulk band gap of $3.2$\,eV, this results in a conduction band offset (CBO) of about 245\,meV from $E_F$, see Figure~\ref{fig:model}~b. We note that the actual STO band gap likely depends on the $V_{\mathrm{O}}$ concentration \cite{Catrou2018,Arras2020}. 
A comparison of the interface formation scenarios as determined from the XPS analysis is sketched in Figure~\ref{fig:model}~a. Qualitatively, the Fermi level shifts away from the CB towards the VB for Fe/STO and FeO/STO, whereas it moves towards or possibly into the CB at the \Fe/STO interface. The experimental parabolic p- and n-type dispersions as observed by ARPES are schematically included in Figure~\ref{fig:model}~b.

\subsection*{Oxygen-vacancy independent emergence of hole- and electron dispersions}
These experimental findings provide a differing and -- according to current knowledge -- unexpected scenario: A hole band dispersion is observed at the Fe/STO interface (see Figure~\ref{fig:HighStats}~b), but concurrently a growth-induced Ti$^{3+,2+}$ $2p$ fraction that is moreover strongly increasing upon x-ray exposure (see Figure~\ref{fig:XPS}~b,\,c). On the other hand, a 2DES emerges at the \Fe/STO interface (see Figure~\ref{fig:HighStats}~d), but without sizeable oxygen vacancy formation during growth or upon x-ray exposure as apparent from the typical Ti $2p$ core-level signatures (see Figure~\ref{fig:XPS}~b,\,c). The FeO/STO case is caught right in-between; a hole-like dispersion is observed (Figure~\ref{fig:HighStats}~c), but oxygen vacancy formation upon growth and x-ray exposure is nearly absent (Figure~\ref{fig:XPS}~b,\,c). 

Our findings reveal that the emergence of either a hole- or electron dispersion at the Fe/STO interface in the present experiment is not critically affected by V$_{\mathrm{O}}$ formation - neither caused by oxygen scavening during growth nor by x-ray illumination. The fact that the Fe/STO interface with such a clear signature of Ti$^{3+,2+}$ displays a hole-type dispersion is unprecedented, though.

From the band structure calculations we infer that the hole-type valence band dispersion observed by ARPES for Fe/STO and FeO/STO interfaces originate from Fe~$3d$ hybridization with Ti~$3d_{xz,yz}$ states in the very first monolayer of the \STO side of the interface. These interfacial hole states may be spatially well separated from a (probably coexisting but) subjacent electron accumulation, as suggested by the large oxygen vacancy concentration found for Fe/\STO. 

%For Fe/STO, we tentatively assume a partial interface structural transformation to a FeO rock salt structure, as FeO(100) can be stabilized onto charge neutral TiO$_2$-terminated \STO(100) surfaces, and Fe can coexist in FeO \cite{Cheng2021}. 

The electron-type 2DES at the \Fe/STO interface likely originates from a downward-shift of the Ti $3d$~$t_{2g}$ CBM and emerging itinerant electronic state at $E_F$. This dispersive quasiparticle state coexists with the in-gap state at -\SI{1.3}{\electronvolt} (see Supporting Information, Figure~\ref{fig:differential}~c), which is typically interpreted as a hallmark feature for the presence of oxygen vacancies in \STO\cite{Walker2015}. The tiny if not absent Ti$^{3+,2+}$ $2p$ core-level signature remains unanticipated, though.

\subsection*{Summary}
To conclude, we provide direct evidence for individual hole- and electron-type valence band dispersions at Fe--\STO heterostructures using momentum-resolved soft x-ray ARPES. Theoretical simulations suggest a model in which the respective band character is set by the Fe oxidation state of an Fe-based overlayer, which tunes hybridization between the Ti- and Fe-derived states across the interface.
A p-type hole band emerges in the band gap of insulating \STO for ultrathin Fe or FeO coverages, respectively, while a 2D electron system is observed for \Fe overlayers. The interface chemistry of the hole-type Fe/\STO interface reveals a strong Ti$^{3+,2+}$~$2p$ contribution, which as of yet had been assigned exclusively to the emergence of 2DES and thus represents a novel case study in the field of 2D oxide interfaces. The results suggest the possibility to deliberately tune between n- or p-type conductivity at \STO-based oxide interfaces, and to exploit the Fe oxidation state to design the resulting band alignment. 
This experimental demonstration of individually emerging 2DHS and 2DES at \STO interfaces may boost novel device concepts and developments, in particular realizing information processing devices with tunable n/p-conductivity - the so far missing link in oxide electronics.

\clearpage

\section*{Experimental Section}

\subsection*{Sample preparation}
Samples were prepared on commercial TiO$_2$-terminated, undoped \STO(001) substrates (from CrysTec) using molecular beam epitaxy (MBE) operated under ultrahigh vacuum (UHV) at a base pressure $p=3\times10^{-11}$\,mbar. The as-received, undoped \STO(STO) substrates were annealed in UHV at \SI{500}{\degreeCelsius} for \SI{30}{\minute}. This surface preparation process leads to the reduction of the TiO$_2$ layers and formation of oxygen vacancies \cite{Fu2005,Walker2015}.

Fe, FeO and \Fe thin films with thicknesses of nominally $1$ or $2$ unit cells (uc) were prepared by e-beam evaporation of metallic Fe at a growth rate of \SI{0.24}{\nano\meter}/min onto STO substrates kept at $T_S=\SI{350}{\degreeCelsius}$. For FeO and \Fe preparation, molecular oxygen was stabilized before and constantly supplied during Fe evaporation at a pressure of $p_{O_2}= 3.3 \times 10^{-7}$\,mbar. For \Fe, the oxygen supply was kept for additional \SI{90}{\second} after the Fe shutter was closed. 

To prevent any (over-)oxidation of the Fe$_x$O$_y$/STO samples, a portable UHV suitcase (Ferrovac) with a base pressure in the $10^{-11}$ mbar range was used to transport the samples in situ to the Swiss Light Source (SLS). All samples were transferred to the measurement chamber without breaking the UHV condition and, hence,  without need for a capping layer.

\subsection*{In-situ sample characterization using LEED and XPS}
The Fe-based overlayer thicknesses were determined during growth by an in situ quartz microbalance. To confirm the crystalline quality of substrates and as-grown films, low-energy electron diffraction (LEED) patterns were recorded in situ using a SPECS ErLEED-1000 on all samples at room temperature, respectively. For bare STO substrates, sharp reflections of the TiO$_2$-terminated surface are obtained, which blur out upon the deposition of 2 uc Fe, FeO and 1uc \Fe, respectively, but without a geometrical change of the reflection pattern.

The chemical composition of the bare \STO substrates and 'as-grown' Fe/STO, FeO/STO and \Fe/STO heterostructures was determined by in situ x-ray photoelectron spectroscopy (XPS) using non-monochromatized Al~K$_\alpha$ radiation ($h\nu = \SI{1486.6}{\electronvolt}$). A setup consisting of a SPECS XR-50 \mbox{x-ray} source and a SPECS Phoibos 100 analyzer was used. 

\subsection*{Quantification of Fe$_x$O$_y$ film thicknesses and stoichiometry}
In addition to the quartz microbalance reading during deposition, the Fe-based film thicknesses were cross-checked by XPS analysis of the Ti\,$2p$ core-level peak attenuation induced by the overlayer. We determine Fe~($\sim$\SI{0.63}{\nano\meter})/STO, FeO~($\sim$\SI{0.83}{\nano\meter})/STO and \Fe~($\sim$\SI{0.96}{\nano\meter})/STO, whereby the unit cell (uc) sizes for Fe, FeO and \Fe are \SI{0.289}{\nano\meter}, \SI{0.43}{\nano\meter} and \SI{0.839}{\nano\meter}, respectively. 

\begin{figure}[h!]
\includegraphics[width=0.7\linewidth]{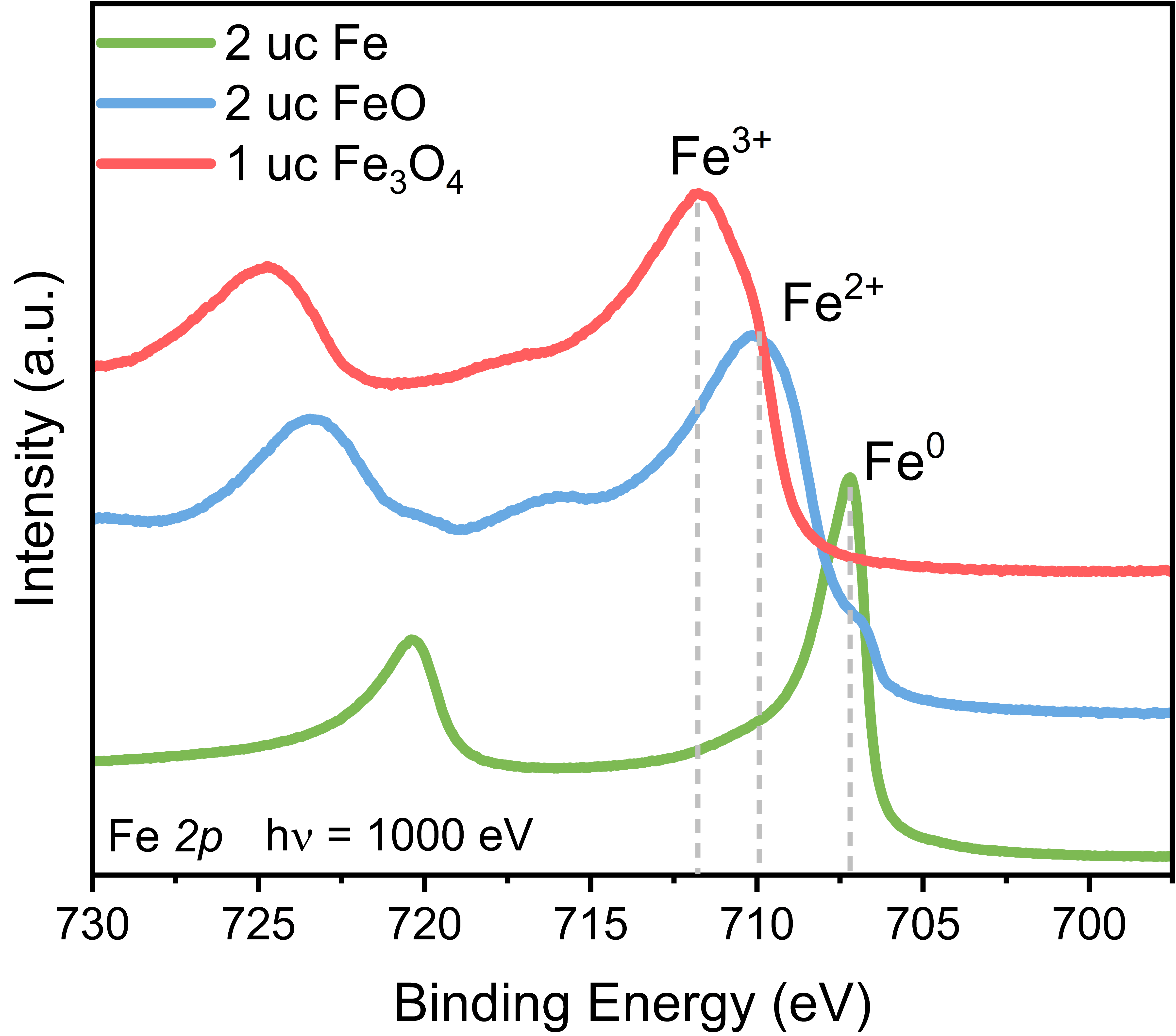}
\caption{XPS spectra of the spin-orbit split Fe 2$p_{3/2}$ and 2$p_{1/2}$ doublets of all 'as-grown' samples. For a quantitative analysis,  a core-level line shape decomposition was performed using reference spectra of Fe metal (Fe$^0$), FeO (Fe$^{2+}$) and magnetite (Fe$^{2+,3+}$)\cite{Hamed2019,Hamed2020}.}
\label{fig:XPSspectra_Fe2p}
\end{figure}

The Fe film stoichiometry of the 'as grown' samples is deduced from the relative positions of the spin-orbit split Fe 2$p_{3/2}$ and 2$p_{1/2}$ doublets, see Supporting Information Figure~\ref{fig:XPSspectra_Fe2p}~a. For Fe$^0$ metal deposited on STO under UHV, they are located at $\sim$\SI{707}{\electronvolt} and $\sim$\SI{720}{\electronvolt}, respectively. For Fe deposited in an O$_2$ atmosphere onto \STO, we determine chemical shifts of the Fe $2p$ doublet towards higher binding energy (BE) indicating an oxidation to Fe$^{2+}$ (FeO) and mixed Fe$^{2+}$/Fe$^{3+}$ states (\Fe).  For a quantitative analysis,  a core-level line shape decomposition was performed using reference spectra of Fe metal (Fe$^0$), FeO (Fe$^{2+}$) and magnetite (Fe$^{2+,3+}$)\cite{Hamed2019,Hamed2020}. 
For FeO/STO the quantitative decomposition of the 'as-grown' Fe $2p$ Lab-XPS spectra reveals a fraction of unoxidized, metallic Fe residuals of about $10\pm0.5$\%. For \Fe/STO, the 'as-grown' Fe $2p$ Lab-XPS spectra reveals a fraction of divalent FeO residuals of about $7\pm0.5$\%.

We compared the evolution of the Fe$^{2+}$ peak shoulder at $\sim$\SI{710}{\electronvolt} for 'as-grown' Fe/STO samples with increasing Fe metal film thickness of 0.1, 0.6\,($\sim2$\,uc) and \SI{10}{\nano\meter}, see Figure~\ref{fig:XPS}~a. Clearly, the Fe$^{2+}$ contribution is suppressed with respect to the Fe$^0$ main peak with increasing Fe metal coverage. This finding confirms that the Fe$^0$ to Fe$^{2+}$ oxidation takes place directly at and originates from the STO interface. The estimated Fe$^{2+}$/Fe$^0$ ratio for the 2 u.c. Fe/STO sample is about 10\% and converts into an approximate thickness of $\sim 0.4$\,ML FeO. 

\begin{figure}[h!]
\includegraphics[width=1\linewidth]{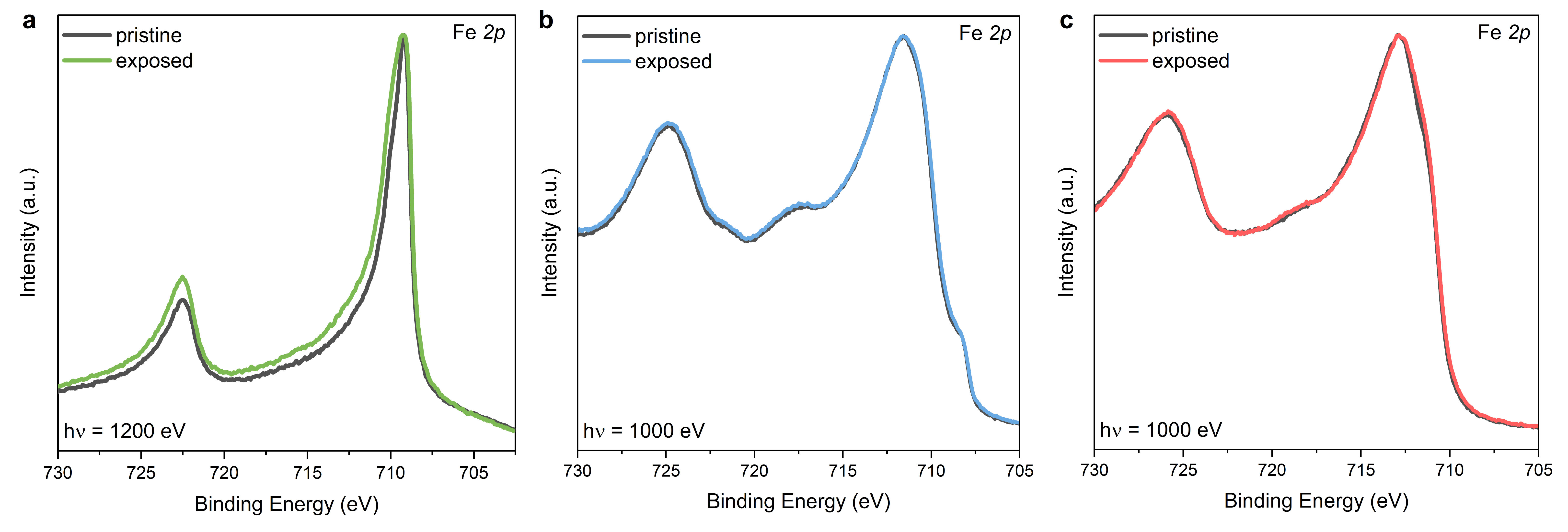}
\caption{Fe $2p$ core-levels of \textbf{a} Fe/STO, \textbf{b} FeO/STO and \textbf{c} Fe$_3$O$_4$/STO in the pristine and x-ray exposed states recorded at the synchrotron. While the x-ray exposure changes the background in Fe/STO, no sizeable change of the line shapes is visible for Fe/\STO, FeO/\STO and \Fe/\STO in both states.}
\label{fig:XPSspectra_Fe2p_TimeDep}
\end{figure}

Finally, we confirm that the line shapes of the Fe $2p_{3/2}$ and $2p_{1/2}$ core-levels for all three Fe--\STO samples remain unchanged during x-ray exposure, although for Fe/STO, there is a change of the background visible in Figure~\ref{fig:XPSspectra_Fe2p_TimeDep} (Supporting Information).

%\textcolor{blue}{Finally, we confirm that the shape of the Fe $2p_{3/2}$ and $2p_{1/2}$ core-levels of the FeO/STO and \Fe/STO samples remains unchanged during x-ray exposure, while Fe/STO further oxidizes depicted in Figure~\ref{fig:XPSspectra_Fe2p_TimeDep} (Supporting Information). This validates the observation at the Ti $2p$ core-levels shown in Figure~\ref{fig:XPS}.}

\subsection*{Angle-resolved photoelectron spectroscopy}
ARPES measurements were performed at the soft x-ray ARPES endstation\cite{Strocov2014} of the Advanced Resonant Spectrocopies (ADRESS) beamline\cite{Strocov2010} of SLS, Paul Scherrer Institute, Switzerland. All experiments were performed with undoped STO substrates. A residual bulk conductivity necessary for PES experiments was facilitated by laser illumination of the samples, which had no measurable influence on the electronic structures. The samples were cooled to $T\sim$\SI{12}{\kelvin} in order to quench the relaxation of k-conservation due to the thermal motion of the atoms\cite{Braun2013}. The combined energy resolution (beamline plus analyzer PHOIBOS-150) was of the order of \SI{40}{meV} for $E(\mathbf{k})$ maps, and the angular resolution of the analyzer was $\sim0.1^{\circ}$. The photon flux was $\sim10^{13}$\, photons/s and focused onto a spot of $30 \times 75$ $\mu$m$^2$ on the sample surface. Variable polarization allowed symmetry analysis of the electron states. All reported ARPES data were collected after more than 1 hour of irradiation time to ensure saturation of x-ray induced processes.

\subsection*{Resonant photoelectron spectroscopy intensity maps}
\begin{figure}[h!]
\includegraphics[width=0.8\linewidth]{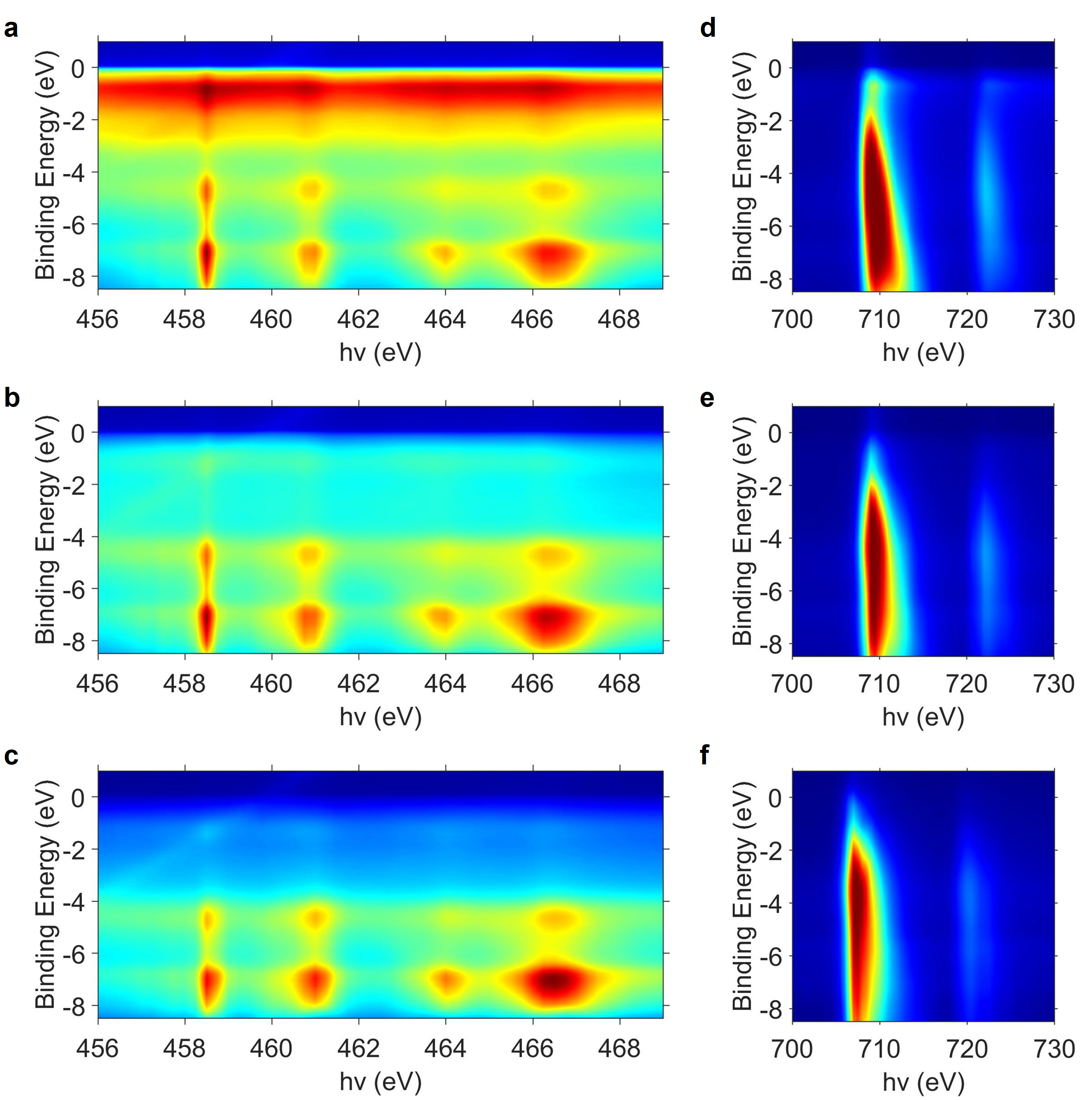}
\caption{Angle-integrated resonant photoemission (resPE) intensity maps across the Ti $L_{3/2}$ and Fe $L_{3/2}$ absorption edges for (a),(d) 2\,uc Fe/STO, (b),(e) 2\,uc FeO/STO and (c),(f) 1\,uc Fe$_3$O$_4$/STO, respectively.}%
\label{fig:ResPESTi}
\end{figure}

A schematics of the resonant photoemission (resPE) process is sketched in Figure~\ref{fig:HighStats}~a. The Ti $L_{3/2}$ edge resonant photoemission intensity maps for (a) 2\,uc Fe/STO, (b) 2\,uc FeO/STO and (c) 1\,uc \Fe/STO interfaces measured at $T=\SI{12}{K}$ and angle-integrated within $\pm k^{yz}$ are depicted in the Supporting Information Figure~\ref{fig:ResPESTi}~a-c. The resonant photoemission intensity maps across the Fe $L_{3/2}$ absorption edges for (d) 2\,uc Fe/STO, (e) 2\,uc FeO/STO and (f) 1\,uc Fe$_3$O$_4$/STO are shown in Supporting Information Figure~\ref{fig:ResPESTi}~d-f.

Analyzing Supporting Information Figure~\ref{fig:ResPESTi}, the spectral weight extending from \SI{-8}{eV} to \SI{-4}{eV} BE is the STO valence band (VB) formed by O$(2p)$ states which hybridize with Ti$(3d)$ and, expectable strongly resonate at the Ti$^{4+}$ $t_{2g}$ and $e_g$ absorption edge. Above the STO VB, from \SI{-3}{eV} BE up to the Fermi level, the otherwise empty band gap of \STO is superimposed by metallic states. For (a) Fe/STO, this energy range is strongly dominated by an incoherent and non-resonating Fe$(3d)$ density of states, which are localized in the Fe overlayer. For (c) \Fe/STO, resonating in-gap states emerge around $\SI{-1.3}{\electronvolt}$.

\begin{figure*}[h!]
\includegraphics[width=1\textwidth]{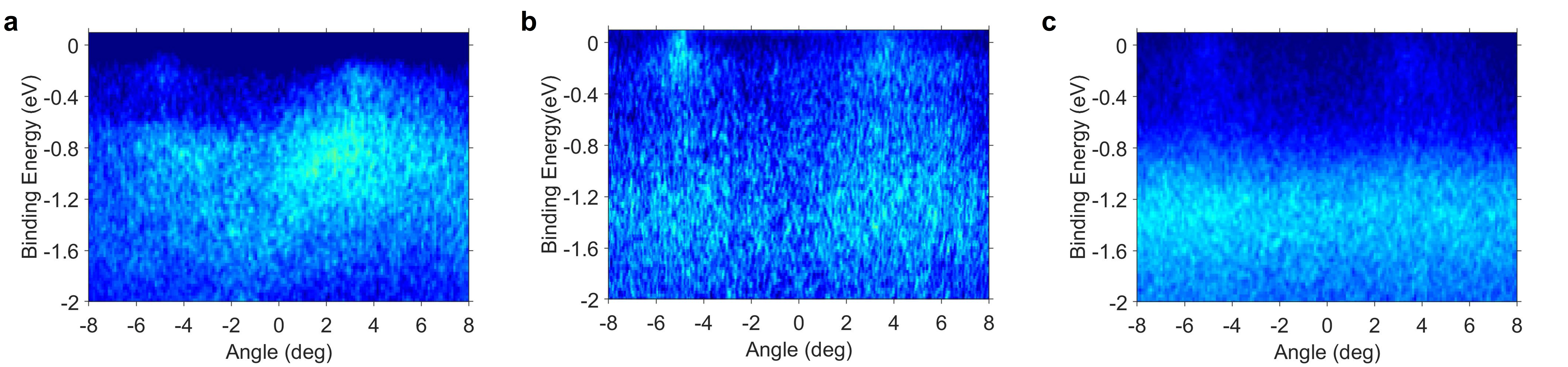}
\caption{Differential spectra of resonant and off-resonant photoemission intensity maps across the Ti $L_{3/2}$ absorption edges for (a) 2\,uc Fe/STO, (b) 2\,uc FeO/STO and (c) 1\,uc Fe$_3$O$_4$/STO.}%
\label{fig:differential}
\end{figure*}

Differential spectra between the on- and off-resonant PES maps across the Ti $L_{3/2}$ absorption edges confirm the accentuated increase of Ti spectral weight at the Ti$^{4+}$ $t_{2g}$ peaks, see Supporting Information Figure~\ref{fig:differential}, and were calculated from photon energies $h\nu=466$\,eV and $h\nu=456$\,eV, respectively. We clearly extract spectral intensity at/below the Fermi level which is resonating with the Ti $L_{3/2}$ absorption edge. In the case of pure STO, spectral intensity at $\SI{-1.3}{\electronvolt}$ is usually related to oxygen vacancy-induced in-gap states resonating at the Ti$^{3+}$ $e_g$ absorption peak\cite{Ohtomo2002}, which due to superposition is not discernable for Fe/STO and FeO/STO.

\subsection*{ARPES data analysis and resPES image processing}
The raw data of the angular resolved resonant valence band structure is a superposition of a $\mathbf{k}$-dispersive spectral fraction with a strong $\mathbf{k}$-integrated background. For data analysis and graphical data processing, the $\mathbf{k}$-integrated background contribution has been simulated by angle integration of the PE intensity. Properly weighted, it has been subtracted from the experimental spectra to uncover the $\mathbf{k}$-dispersive fraction. These dispersive structures displayed in Figure~\ref{fig:HighStats}~b-d have been extracted by taking the maximum curvature of the smoothed photoemission intensity along the BE direction\cite{Zhang2011}. This data analysis used MATLAB-based software ARPESView for viewing and processing of the ARPES data\cite{ARPESview}.

\subsection*{Band structure calculations} 
Electronic band structures were calculated using a first-principles approach building up on a mixed-basis pseudopotential scheme based\cite{elsaesser90,lechermann02} on Kohn-Sham density functional theory (DFT) supplemented by an additional treatment of electronic correlations. Norm-conserving pseudopotentials and a mixed basis consisting of localized functions and plane waves are utilized. Localized functions are introduced for Ti$(3d)$, Fe$(3p,3d)$ and O$(2s,2p)$. A plane-wave cutoff energy of $E_{\rm cut}=11$\,Ry and a $7\times 7\times 1$ $k$-point mesh are used for the final supercell calculations.

To reveal the spectral data, the DFT+sic~\cite{vogel96,koerner10} methodology is utilized, which is suited to include electronic correlations beyond conventional DFT in an efficient way for large-scale supercell computation. It corresponds to overall applying the local density approximation (LDA) but employing here an oxygen pseudopotential which is modified by the self-interaction correction (SIC)~\cite{perdew81} in order to address local Coulomb interactions. While the O$(2s)$ orbital is by default fully corrected with a weight factor $w_{2s}=1.0$, the reasonable choice~\cite{koerner10,lechermann19} $w_{2p}=0.8$ is utilized for the O$(2p)$ orbitals. A further screening parameter $\alpha$ for this SIC pseudopotential is also chosen as $\alpha=0.8$, such that the SIC inclusion on oxygen asks for one additional parameter in the overall computational scheme. Note that the description of electronic correlations originating here only from the oxygen sites may be a good approximation for weakly-to-moderately correlated systems, because of the hybridization between transition-metal and oxygen sites. For instance, such a plain DFT+sic approach to the moderately-correlated metal SrVO$_3$ results in a low-energy band renormalization of $Z\sim 0.6$, in very good agreement with experiment. In the present context, application of this DFT+sic setting to bulk STO leads to a band gap $\Delta=2.8$\,eV close to the experimental value, while conventional DFT usually yields $\Delta\sim 2$\,eV.

\begin{figure*}[h!]
\includegraphics[width=0.65\textwidth]{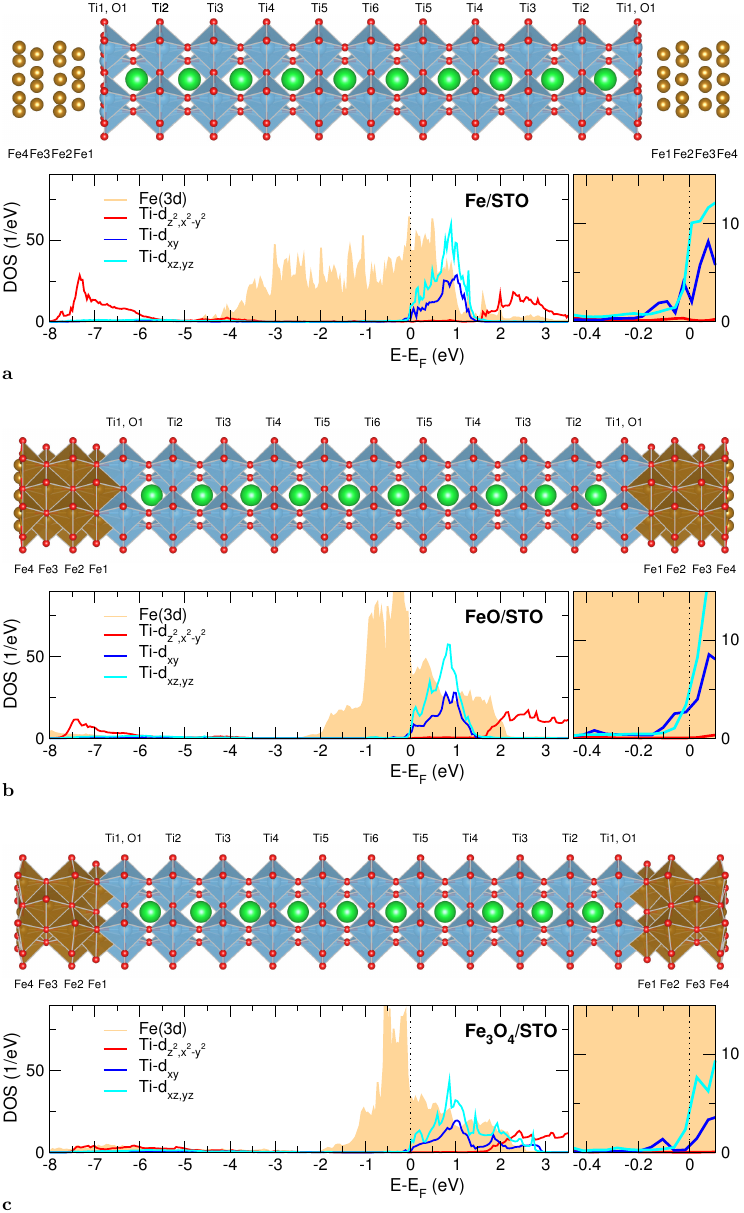}
\caption{Supercell data for (a) Fe/STO, (b) FeO/STO and (c) Fe$_3$O$_4$/STO. Top: Symmetric slab geometry for each structural case, involving Fe (orange), Sr (green), Ti (blue) and O (red). The $c$-axis points along the horizontal axis. Bottom: Projected density of states (DOS) for Fe$(3d)$ and Ti$(3d)$ as obtained from the DFT+sic calculations. Integration over all layers is understood. The outermost right panel displays a blowup of the low-energy region, respectively.}
\label{fig:struc}
\end{figure*}

Tetragonal supercells in slab architecture based on a $1\times 1$ in-plane motif 
were constructed for the three structural cases of Fe/STO (69 atoms), FeO/STO (85 atoms) and \Fe/STO (81 atoms) as shown in the respective upper part of Figure~\ref{fig:struc}. 
Symmetric slabs with in total 11 TiO$_2$ layers for the STO part are chosen. There are thus 6 layer-inequivalent Ti sites, with the mirror plane passing through the Ti6O$_2$ layer. On top of each STO surface, four Fe layers with and without oxygen coordination are added, whereby the first Fe1 layer is placed on top of the O1 positions of STO, respectively. This positioning is energetically favored compared to Fe on top of Ti. 
For Fe/STO case, a bcc-Fe coordination is realized, while FeO on STO amounts to a three-dimensionally alternating, i.e. simple-cubic Fe,O lattice. Concerning \Fe/STO, for computational reasons we did not employ the bulk \Fe structure with differentiation of octahedral and tetrahedral Fe sites. Instead a structural approximant, building up on iron vacancies in the FeO layers tailored to the \Fe stoichiometry is chosen. Note that the true \Fe/STO structure may not be fully ordered in the \Fe-bulk sense, and therefore our approximant is believed to be a meaningful description to evaluate the key features of the \Fe/STO electronic structure.
The experimental STO lattice parameter $a=3.905$\,\AA\, is utilized for all structures and structural relaxation within the generalized-gradient approximation is performed to optimize the atomic positions, resulting in different layer-to-layer distances and local distortions. Note that the Ti5,Ti6 (and connected Sr,O sites) positions are kept fixed to the experimental STO lattice positions, in order install the 'STO bulk limit' within the interior of the slabs. The relaxed vacuum region between periodically repeated slabs along the $c$ direction amounts to a safe distance of 22-25\,\AA\, depending on the structural case. The nearest-neighbor (NN) Fe1-O1 distance reads $d_{\rm Fe1-O1}=1.942$\,\AA\, for Fe/STO, $d_{\rm Fe1-O1}=1.926$\,\AA\, for FeO/STO and $d_{\rm Fe1-O1}=1.935$\,\AA\, for \Fe/STO. The NN Fe1-Fe1 distance differs only slightly from the value $d_{\rm Fe1-Fe1}=2.76$\,\AA\, throughout the different structural cases.

The lower part of Figure~\ref{fig:struc} shows correspondingly the projected Fe$(3d)$ and Ti $3d$ density of states (DOS). Non-surprisingly, the Fe $3d$ DOS extends over a wide energy range into the original STO band gap. The largest extension for Fe/STO case is easily understood from the highly itinerant bcc-Fe overlayers. The more correlated FeO and \Fe overlayers result in an effective Fe$(3d)$ bandwidth reduction. The Ti$(3d)$ low-energy states of exclusive $t_{2g}=\{xz,yz,xy\}$ kind are in all cases dissolved in the Fe $3d$ background. Notable Ti $e_g=\{z^2,x^2-y^2\}$ weight, most prone to O $2p$ hybridization, resides at higher energies up and below the Fermi level. Oxygen vacancies are effective in bringing such $e_g$ weight to lower energies (not shown).
Note that the dominantly dispersing Ti $d_{xz,yz}$ character shown in Figure~\ref{fig:DFT} is associated with the Ti1 sites, respectively. The notable Ti $d_{xy}$ low-energy character, best visible in the outermost right part of Figure~\ref{fig:struc}, is associated with contributions from the deeper Ti layers, mostly Ti2-Ti4.

\subsection*{Supporting Information}
Supporting Information is available from the Wiley Online Library or from the author.

\subsection*{Acknowledgements}
We thank T. Szyjka for help with the XPS analysis. F.L. acknowledges support from the European XFEL and the Center for Computational Quantum Physics of the Flatiron Institute under the Simons Award ID 825141. Computations were performed at the JUWELS Cluster of the J\"ulich Supercomputing Centre (JSC) under projects hhh08 and miqs. F.A. thanks the Swiss National Science Foundation for financial support within the grant 200020B-188709. M.M. acknowledges funding by the Deutsche Forschungsgemeinschaft (DFG, German Research Foundation), SFB 1432 (Project B03) with Project-ID 425217212.

\subsection*{Author contributions}
M.M. conceived the experiment. P.R. performed preliminary experiments. P.M.D. deposited and characterized the samples. P.M.D., P.R., M.M., F.A. and V.N.S. performed the ARPES experiments. P.M.D. performed the ARPES analysis under supervision of V.N.S.. P.M.D. performed the XPS- and band alignment analysis under supervision of M.M. and L.B..  P.M.D., L.B., V.N.S. and M.M. interpreted the experimental data with help of all co-authors. F.L. conceived the theory part and performed the simulations. M.M. and F.L. wrote the manuscript with help of all co-authors.

\subsection*{Conflict of interests}
The authors declare no competing interests. 

\subsection*{Keywords}
2D electron gases, 2D hole gases, oxide-based electronics, oxide interfaces 

\clearpage

%%%%%%%%%%%%%

% Figures for Supporting Information
\section*{Supporting Information}

\subsection*{XPS spectra of the spin-orbit split Fe 2$p_{3/2}$ and 2$p_{1/2}$ multiplet}

\clearpage

\subsection*{Time-dependent XPS spectra of the Fe 2$p_{3/2}$ and 2$p_{1/2}$ core-levels}

\clearpage

\subsection*{Angle-integrated PES maps across the Ti $L_{3/2}$ and Fe $L_{3/2}$ absorption edges}

\clearpage

\subsection*{Differential spectra at the Ti $L_{3/2}$ absorption edges}

\clearpage

\subsection*{Supercell calculations}

\clearpage
\nolinenumbers
%%%%%%%%%
% Bib-File für Methods & Extended Data
\bibliography{holegas.bib}
%%%%%%%%%

\end{document}